\documentclass[twocolumn, trackchanges]{aastex7}
\usepackage{CJKutf8}
\usepackage{amsmath}  
\usepackage{xcolor} 
\newcommand{\BALQSO}{BAL quasar}
\newcommand{\BALQSOs}{BAL quasars}
\newcommand{\NONBAL}{non-BAL quasar}
\newcommand{\NONBALs}{non-BAL quasars}
\newcommand{\DR}{SDSS DR16Q}
\newcommand{\Med}{\texttt{BALNet}}
\newcommand{\PROB}{PIXEL\_PROB}   

\begin{document}
\begin{CJK}{UTF8}{gbsn}

\title{BALNet: Deep Learning-Based Detection and Measurement of Broad Absorption Lines in Quasar Spectra}

\author{Yangyang Li}
\affiliation{Shanghai Key Lab for Astrophysics, Shanghai Normal University, Shanghai 200234, China}
\email{liyangyang11223@163.com}  

\author[0009-0009-1617-8747]{Zhijian Luo}
\affiliation{Shanghai Key Lab for Astrophysics, Shanghai Normal University, Shanghai 200234, China}
\email[show]{zjluo@shnu.edu.cn}  

\author[0000-0001-8485-2814]{Shaohua Zhang}
\affiliation{Shanghai Key Lab for Astrophysics, Shanghai Normal University, Shanghai 200234, China}
\email[show]{zhangshaohua@shnu.edu.cn}  

\author{Du Wang}
\affiliation{Shanghai Starriver Bilingual School, Shanghai 200233, China}
\email{du_wang2024@163.com}

\author{Jianzhen Chen}
\affiliation{Shanghai Key Lab for Astrophysics, Shanghai Normal University, Shanghai 200234, China}
\email{jzchen@shnu.edu.cn}

\author[0000-0002-2326-0476]{Zhu Chen}
\affiliation{Shanghai Key Lab for Astrophysics, Shanghai Normal University, Shanghai 200234, China}
\email{zhuchen@shnu.edu.cn}

\author[0000-0001-8244-1229]{Hubing Xiao}
\affiliation{Shanghai Key Lab for Astrophysics, Shanghai Normal University, Shanghai 200234, China}
\email{hubing.xiao@shnu.edu.cn}

\author{Chenggang Shu}
\affiliation{Shanghai Key Lab for Astrophysics, Shanghai Normal University, Shanghai 200234, China}
\email{cgshu@shao.ac.cn}

\begin{abstract}

Broad absorption line (BAL) quasars serve as critical probes for understanding active galactic nucleus (AGN) outflows, black hole accretion, and cosmic evolution. To address the limitations of manual classification in large-scale spectroscopic surveys - where the number of quasar spectra is growing exponentially - we propose \Med, a deep learning approach consisting of a one-dimensional convolutional neural network (1D-CNN) and bidirectional long short-term memory (Bi-LSTM) networks to automatically detect BAL troughs in quasar spectra. \Med\ enables both the identification of \BALQSOs\ and the measurement of their BAL troughs. We construct a simulated dataset for training and testing by combining \NONBAL\ spectra and BAL troughs, both derived from SDSS DR16 observations. Experimental results in the testing set show that: (1) BAL trough detection  achieves 83.0\% completeness, 90.7\% purity, and an F1-score of 86.7\%; (2) \BALQSO\ classification  achieves 90.8\% completeness and 94.4\% purity; (3) the predicted BAL velocities agree closely with simulated ground truth labels, confirming \Med's robustness and accuracy. When applied to the SDSS DR16 data within the redshift range  $1.5<z<5.7$, at least one BAL trough is detected in 20.4\% of spectra. Notably, more than a quarter of these are newly identified sources with significant absorption,  8.8\% correspond to redshifted systems, and some narrow/weak absorption features were missed. \Med\ greatly improves the efficiency of large-scale BAL trough detection and enables more effective scientific analysis of quasar spectra.
\end{abstract}



\section{Introduction} \label{section:introduction}

Quasars are luminous active galactic nuclei (AGNs) powered by accreting supermassive black holes (SMBHs), and their powerful outflows are generally belived to play a crucial role in galaxy evolution.  These outflows carry away huge amounts of material, kinetic energy, and angular momentum from the nuclear region, they may heat or expel interstellar gas, suppress star formation in the host galaxy, and regulate SMBH growth  (e.g., \citealt{2006_HopkinsHernquist_ApJS..166....1H,2010_HopkinsElvis_MNRAS.401....7H}).
The most prominent signature of quasar outflows is the presence of blueshifted (up to $\rm \sim 0.2c$) and broad (at least 2000~ km/s) absorption line (BAL) troughs, which appear in both high-ionization species (e.g., \ion{C}{4}, \ion{Si}{4}, and \ion{N}{5}; \citealt{1991_Weymann_ApJ...373...23W,2001_Brotherton_ApJ...546..775B,2003_Reichard_AJ_125}), low-ionization species (e.g., \ion{Mg}{2}, \ion{Al}{3}, and \ion{He}{1}*; \citealt{2002_Tolea_ApJ_578,2003_Hewett_AJ....125.1784H,2010_Zhang_ApJ_714,2011_Zhang_RAA11_1163Z,2015_Liu_ApJS_217}), and even in the excited of states \ion{Fe}{2} and/or \ion{Fe}{3} \citep{1987_Hazard_ApJ...323..263H,1997_Becker_ApJ...479L..93B,2012_Vivek_MNRAS.423.2879V,2015_Zhang_ApJ...803...58Z}.
Statistical studies show that approximately 10–20\% of ultraviolet- and optically selected
quasars exhibit BAL troughs \citep{1990_Foltz_BAAS...22..806F,2006_Trump_ApJS..165....1T,2007_Ganguly_ApJ_665,2009_Gibson_ApJ...692..758G}. It is likely that this fraction depends on the orientation and inner structure of the AGN \citep{2002_Tolea_ApJ_578,2003_Hewett_AJ....125.1784H}, and may also reflect evolutionary phases of quasar activity \citep{1988_Sanders_ApJ...325...74S,1993_Hamann_ApJ...418...11H,1993_Voit_ApJ...413...95V}. Moreover, the incidence, variability, and strength of BAL troughs from various ionic species provide crucial diagnostics for the physical and dynamical properties of outflowing gases, as well as valuable insights into their origin and acceleration mechanisms (e.g., \citealt{2013_Filiz_ApJ...777..168F,2014_Vivek_MNRAS.440..799V,2015_HeZhiCheng_MNRAS.454.3962H,2016_ShiXiheng_ApJ...819...99S,2018_Hamann_MNRAS.476..943H,2022_Chen_NatAs...6..339C}). Therefore, the accurate identification and measurement of BAL troughs in quasar spectra constitute a fundamental step in the study of quasar outflows.

Traditional methods for identifying BAL troughs include quasar composite spectrum fitting combined with visual inspection \citep{2003_Reichard_AJ_125,2006_Trump_ApJS..165....1T}, multi-component spectral decomposition \citep{2002_Tolea_ApJ_578,2009_Gibson_ApJ...692..758G,2010_Zhang_ApJ_714}, principal component analysis (PCA; \citealt{1998_Glazebrook_PCA_ApJ...492...98G}), 
non-negative matrix factorisation (NMF; \citealt{2008_ALLEN_NMF_AIPC.1082...85A}), 
and quasar spectrum pair-matching technique \citep{2014_Zhang_ApJ_786,2015_Liu_ApJS_217}.
These approaches involve fitting each individual quasar spectrum to reconstruct the intrinsic (unabsorbed) quasar spectrum, and then identifying BAL troughs by comparing the observed and intrinsic spectra. 
While effective, these methods are time-intensive and not easily scalable to the large and rapidly growing spectroscopic datasets produced by modern astronomical surveys.
For example, the Sloan Digital Sky Survey (SDSS; \citealt{2000_York_AJ....120.1579Y}) DR16 quasar catalog (DR16Q; \citealt{2020_Lyke_ApJS..250....8L}) has provided 750,414 optical quasar spectra. The Large Sky Area Multi-Object Fiber Spectroscopic Telescope (LAMOST; \citealt{2012_Gui_RAA....12.1197C}) has also identified a total of 56,175 quasars \citep{2023_JinJunJie_ApJS..265...25J}. The ongoing larger  survey, the Dark Energy Spectroscopic Instrument (DESI; \citealt{2019_Dey_AJ....157..168D}), aims to quadruple the known quasar sample by obtaining spectra of nearly 3 million quasars \citep{2023_EChaussidon_AAS_944}. The rapid increase in data volume demands the development of more efficient and scalable methods for identifying BAL troughs. 

With the advancement of computer technology, machine learning and artificial intelligence methods have been increasingly applied to the classification and feature recognition of astronomical spectral data. Commonly used techniques include K-Nearest Neighbors (KNN; e.g., \citealt{2007_ZhangZhou_PatRe..40.2038Z,2011_fushiki_estimation,2020_sookmee_17th}), Support Vector Machines (SVM; e.g., \citealt{2015_Liu_SVM_RAA....15.1137L,2020_Barrientos_ASPC..522..385B}), Decision Trees (DT; e.g., \citealt{1996_quinlan_DTlearning,2014_Czajkowski_AIM_61}), and Artificial Neural Networks (ANN; e.g., \citealt{2017WangGuoLuo_MNRAS_465,2018_Busca_arXiv180809955B,2022_Rastegarnia_FNet}). 
More recently, \cite{2023_Yang_MNRAS.518.5904Y} demonstrated that convolutional neural networks (CNNs) significantly outperform traditional methods in classifying observed astronomical spectra. 
\cite{2018_Busca_arXiv180809955B} proposed a deep CNN model, \texttt{QuasarNET}, designed for redshift estimation and object classification, including the identification of \BALQSOs. In addition, \cite{2019_Guo_ApJ...879...72G} developed a PCA-enhanced CNN framework for classifying \BALQSOs.
\cite{2024_Moradi_MNRAS.533.1976M} introduced an enhanced ResNet-based CNN model,
\texttt{FNet II}, capable of processing spectra through automated feature extraction, thereby eliminating the need for manual identification of spectral lines.
Comparative studies by \cite{2024_Kao_PASJ...76..653K} and \cite{2025_Pang_AS} have further established the superiority of deep learning approaches, particularly CNN architectures, over traditional machine learning methods like extreme gradient boosting (XGBoost) for \BALQSO\ identification.

Although existing methods have achieved impressive performance in classifying \BALQSOs\ — with extremely  high detection completeness exceeding 97\% (e.g., \citealt{2018_Busca_arXiv180809955B,2019_Guo_ApJ...879...72G,2024_Moradi_MNRAS.533.1976M,2024_Kao_PASJ...76..653K,2025_Pang_AS}) — several key challenges remain unresolved.
First, most current approaches focus primarily on distinguishing \BALQSOs\ from general quasar populations, but lack the ability to quantitatively characterize BAL troughs, such as their velocity and velocity structure.
Second, the training datasets are mainly constructed from previously confirmed \BALQSOs, which tend to be biased toward strong \ion{C}{4} BAL troughs with large blueshifted velocities.
Although the high completeness and purity of these datasets are notable, they do not truly reflect the performance on real \BALQSOs. 
This leads to a selection bias that limits model sensitivity to shallower troughs or those superimposed on \ion{C}{4} emission lines, thereby reducing the completeness and generalizability of the classification.
In addition, the important subclass of redshifted BALs has been largely overlooked.
Previous studies have shown that they trace inflowing gases in the nuclear region of AGNs \citep{2017_Shi_ApJ...843L..14S, 2017_Zhang_ApJ...839..101Z, 2019_Zhou_Natur.573...83Z}, providing key insights into SMBH accretion physics.
Therefore, it is necessary to address these limitations from both methodological and training dataset construction perspectives.

In this paper, we introduce \Med, a new automated deep learning framework that integrates a one-dimensional convolutional neural network (1D-CNN) with bidirectional long short-term memory (Bi-LSTM) networks to accurately detect and characterize \ion{C}{4} BAL trough in quasar spectra. Unlike previous methods, \Med\ not only classifies \BALQSOs\ but also directly measures the kinematic properties (e.g., velocities) of \ion{C}{4} BAL troughs. For model training and evaluation, we generated a large set of simulated spectra with diverse \ion{C}{4} BAL trough profiles, derived from the \DR\ dataset.

The remainder of this paper is organized as follows: 
Section \ref{sec:Data} describes the construction of mock spectra used for model training and testing. 
Section \ref{sec:deepLearning} details the architecture of the proposed \Med\ framework and outlines its training procedure.
Section \ref{sec:resultanddiscussion} presents the evaluation results of \Med\ on simulated test data and discusses its performance.
Section \ref{sec: resultDr16} then applies \Med\ to the \DR\ dataset and analyzes its performance.
Finally, Section \ref{sec:conclusions} summarizes the main conclusions of this study and outlines future directions.

\section{Mock Data}\label{sec:Data}

The performance of deep learning models depends heavily on both the quantity and quality of the training dataset. A large-scale and accurately labeled training sample is a key factor in improving model performance. 
It enables deep learning models to effectively learn diverse features, thereby significantly enhancing their generalization ability and overall prediction accuracy.
The aim  in this study is how to simultaneously identify \BALQSOs\ and measure the BAL troughs. 
All existing labeled quasar datasets provide only conventional BAL classifications (i.e., whether a quasar is a \BALQSO) and lack detailed annotations of specific BAL velocity structures. As a result, such datasets are inadequate for our research objectives.

To address this, we construct a more comprehensive and accurately labeled mock dataset of quasar spectra for training and evaluating the \Med\ model. This mock dataset is based on the \DR\ dataset, limited to quasars with redshifts ranging from 1.5 to 5.7 to ensure that potential \ion{C}{4} BAL troughs fall within the SDSS spectrograph’s wavelength coverage.
The \DR\ catalog provides the BI\_CIV and AI\_CIV parameters, which are used both to distinguish \BALQSOs\ from \NONBALs\ and to quantify the strength of \ion{C}{4} BALs,  we then obtain 23,994 \BALQSOs\ with BI\_CIV $>$ 0 and 313,739 \NONBALs\ with BI\_CIV $=$ 0 and AI\_CIV $=$ 0.
In the construction of the mock dataset, spectra of \NONBALs\ are randomly sampled, while \ion{C}{4} BAL troughs are extracted directly from real absorption troughs in observed \BALQSOs. The procedures for \ion{C}{4} BAL trough extraction and mock spectrum generation are described in detail in the following two subsections.
Before analysis, all spectra are corrected for Galactic extinction using the extinction map of \citet{1998_Schlegel_ApJ} and the reddening curve of \citet{1999_Fitzpatrick_PASP}, and then shifted to the rest frame using the primary redshift from the \DR. 

\subsection{BAL Trough Extraction} \label{ssec:DataPreparation}

We employ the quasar spectrum pair-matching method (\citealt{2014_Zhang_ApJ_786,2015_Liu_ApJS_217}) to extract the \ion{C}{4} BAL troughs from the spectra of the \DR\ \BALQSOs. 
The core idea, similar to other traditional approaches, is to construct an unabsorbed model spectrum that closely matches the absorption-free regions of a given \BALQSO\ candidate. 
This model is then compared with the observed spectrum to identify potential BAL troughs.

The specific implementation consists of the following three steps:
(1) Randomly select 300 \NONBAL\ spectra with redshifts close to that of the \BALQSO\ candidate under analysis to form a template library. Each template spectrum is smoothed through two iterations of B-spline fitting to remove narrow absorption lines.
(2) Each template is reddened using the SMC extinction law and fitted to the absorption-free regions of the given \BALQSO\ candidate. The scaling factor and the color excess $E(B-V)$ are optimized by the minimizing $\chi^2$ between the model and the observed spectrum.
To improve the fitting accuracy and robustness, we restrict the analysis to a rest-frame wavelength range of 1300 - 1700 \AA, which ensures coverage of the \ion{C}{4} BAL trough while minimizing contamination from unrelated spectral regions.
(3) All model spectra are ranked in ascending order of total $\chi^2$, and the top forty best-fitting models are selected. For each of them, an additional emission-line $\chi^2$ metric, $\chi^2_{\rm EL}$, is calculated over the 1450 - 1650 \AA\ range, excluding pixels affected by BAL troughs. This metric quantifies the similarity between the model and the observed spectrum in the unabsorbed \ion{C}{4} emission-line region. Finally, the composite of the ten models with the lowest $\chi^2_{\rm EL}$ values is adopted as the final unabsorbed model spectrum.
This method exhibits strong detection performance for shallow and weak BAL features, and achieves high accuracy in measuring absorption velocities.

\begin{figure}[hpbt]
	\centering
	\includegraphics[width=0.8\linewidth]{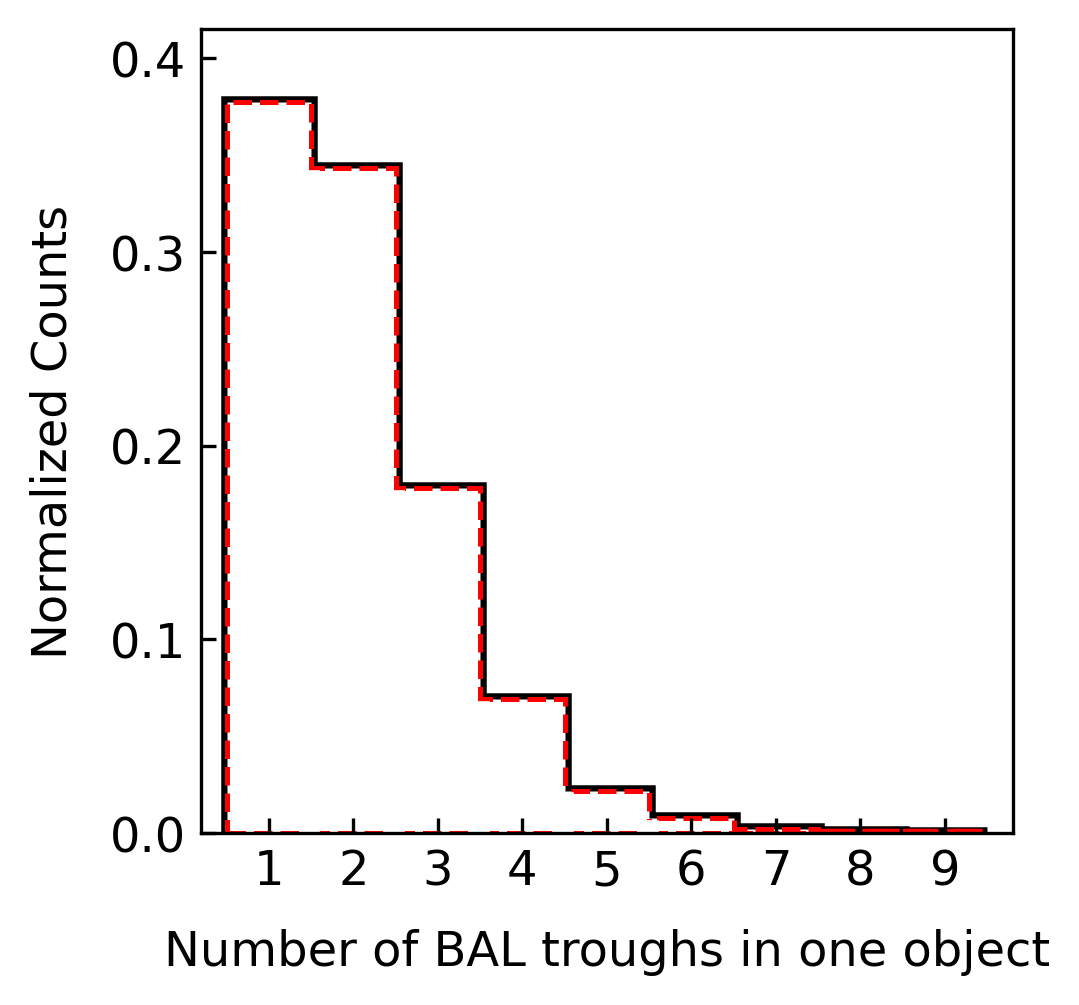}
	\caption{Distributions of the number of \ion{C}{4} BAL troughs per spectrum for 23,107 observed \BALQSOs\  (black solid line) and for 100,000 simulated \BALQSOs\ (red dashed line).}
	\label{fig:Obser} 
\end{figure}

We systematically search for BAL troughs in the normalized spectra of \BALQSO\ candidates. In velocity space, the search is conducted over the range from \( v_l = 10{,}000\, \)~ km/s to \( v_u = -29{,}000\, \)~ km/s, where positive velocities indicate redshifted absorption and negative velocities indicate blueshifted absorption. 
Only features with a continuous absorption over a velocity interval greater than \( 1{,}000\, \)~ km/s for a depth of at least 10\% (normalized fluxes smaller than 0.9) are considered valid BAL troughs.
After screening 23,994 DR16Q \BALQSO\ candidates, we successfully detected 47,267 BAL troughs in 23,107 sources. In Figure~\ref{fig:Obser}, we present the number distribution of BAL troughs in these observed \BALQSOs\ (black solid line). 
The results show that most \BALQSOs\ contain one to three BAL troughs, while a small fraction exhibit four or more, and in rare cases, up to nine troughs. All identified BAL troughs constitute the BAL pattern library, which is used to construct simulated spectra for subsequent model training. 

\subsection{Data Construction} \label{ssec:simulatedSet}
\begin{figure*}[hpbt]
	\centering
	\includegraphics[width=\linewidth]{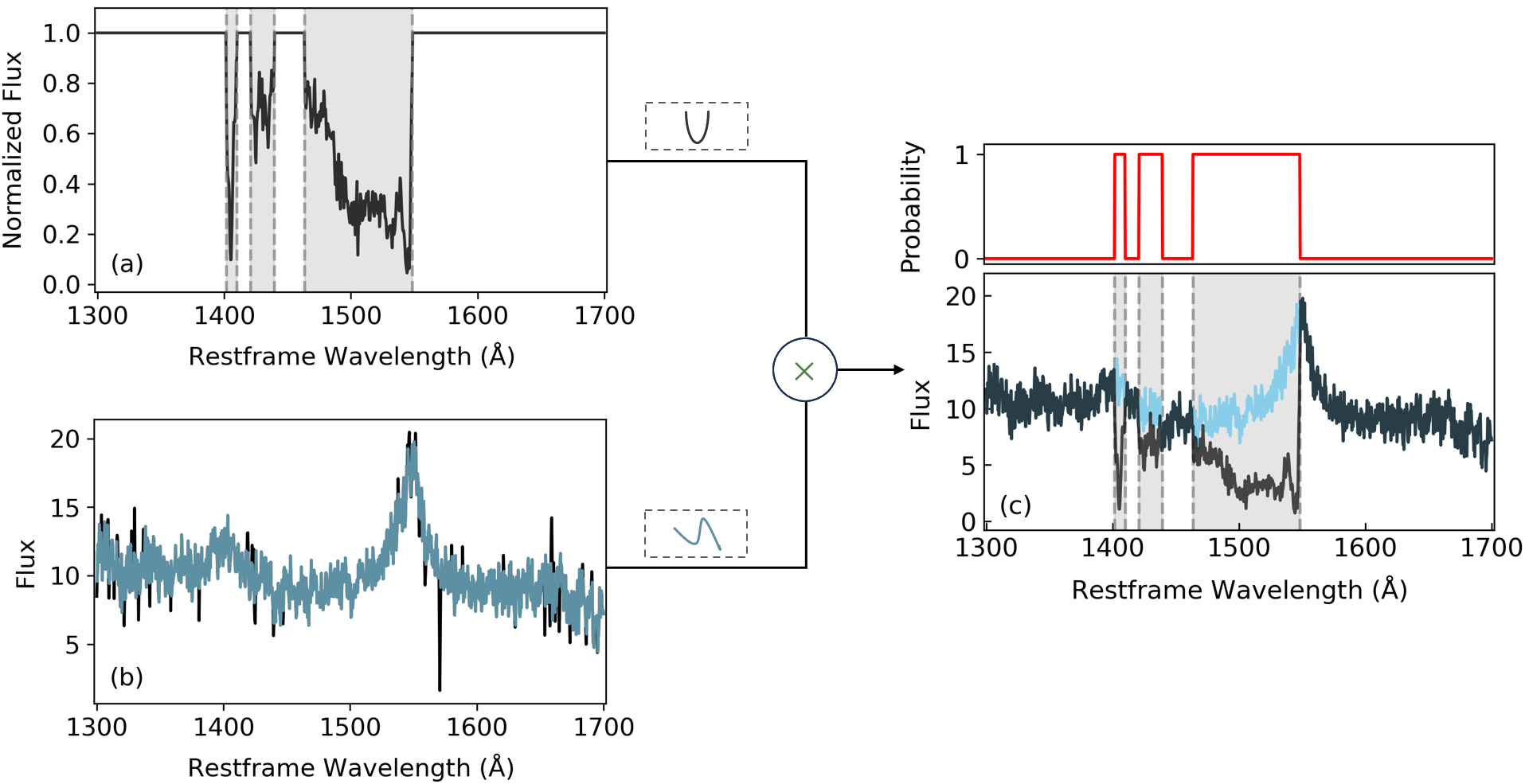}
	\caption{Flowchart illustrating the procedure for constructing simulated spectra. 
    Panel (a) presents a normalized BAL pattern spectrum (black), with the gray-shaded regions marking three randomly selected BAL troughs. 
    Panel (b) displays a real, unabsorbed spectrum (black) and its smoothed version (blue). 
    Panel (c) shows the final simulated spectrum (black) in the bottom part, and the corresponding label vector (red) in the upper part, where BAL regions are labeled as 1 and non-BAL regions as 0. }\label{fig:ConstructSample}
\end{figure*}

In this subsection, we construct a large-scale mock dataset of simulated \BALQSO\ spectra by combining the high-quality BAL pattern library with \NONBAL\ samples derived from real observations. These simulated spectra not only encompass a wide variety of BAL trough types, but also include detailed annotations of the corresponding velocity parameters, thereby satisfying the requirements for simultaneous BAL classification and velocity feature extraction.
Figure \ref{fig:ConstructSample} illustrates the procedure for constructing a representative simulated \BALQSO\ spectrum. The construction process strictly follows the steps outlined below.

Firstly, we randomly select $n$ BAL troughs from the BAL pattern library constructed in the previous subsection, where $n$ is a positive integer (e.g., 1, 2, 3, etc.). 
These patterns are then placed at random, non-overlapping positions within the velocity range from
\(10{,}000\) to \(-29{,}000\, \)~ km/s to generate the normalized BAL spectrum ($NS_{BALs}(\lambda)$).
In these spectra, wavelength regions without BAL patterns have a flux value of 1. Figure \ref{fig:ConstructSample} (a) shows a typical normalized BAL spectrum, with gray-shaded areas representing 3 BAL troughs.

Secondly, background spectrum smoothing is performed. We randomly selected an original spectrum from the \NONBAL\ sample 
and applied a three-point smoothing method.  This process aims to reduce the impact of noise and unresolvable absorption lines, thereby minimizing their interference with subsequent model analysis. 
Figure \ref{fig:ConstructSample} (b) illustrates an example: the raw spectrum of the SDSS object is shown in black, while its smoothed counterpart, denoted as $f_{non-BAL}(\lambda)$, is shown in blue.

Finally,  the background spectrum ($f_{non-BAL}(\lambda)$) is combined with the normalized BAL spectrum ($NS_{BALs}(\lambda)$) to generate the simulated \BALQSO\ spectrum ($f_{simulated}(\lambda)$), as shown by the black line in Figure \ref{fig:ConstructSample} (c). The formula is expressed as:
\begin{equation}
		f_{simulated}(\lambda) = NS_{BALs}(\lambda) \times f_{non-BAL}(\lambda).
\end{equation}    
To generate the final simulated \BALQSO\ spectrum data set, we performed the above procedure 100,000 times. The number distribution of BAL troughs in simulated spectra is consistent with the observational data (see Figure \ref{fig:Obser}), thereby ensuring the high credibility of the simulation results. Additionally, to ensure balance and stability in the training dataset, we randomly selected 100,000 spectra from \NONBAL\ sample and combined them with all simulated \BALQSO\ spectra. The background spectra of the simulated \BALQSOs\ and the \NONBAL\ spectra were not subjected to any signal-to-noise cuts, ensuring that their signal-to-noise distributions are consistent with those of SDSS DR16Q. Through this integration, we created a comprehensive dataset containing a total of 200,000 spectra. This dataset was used to train and test our \Med\ model.

It should be noted that the training set spectra ultimately constructed encompass various manifestations of BAL troughs, such as blueshifted, redshifted, and multiple BAL troughs. Meanwhile, all spectra in the training set have been processed through interpolation and are evenly distributed across 1165 equally spaced wavelength points within the range of 1300 to 1700 \AA. In our dataset, in addition to the spectral data, label data for each spectrum are also required. The most straightforward approach would be to label each wavelength point of the spectrum as to whether it belongs to a BAL trough, for example, using 1 to indicate the presence of a BAL trough and 0 for its absence. However, this method would result in overly complex label data and excessively high computational costs. To avoid this issue, we defined coarser interval points within the same wavelength range as the simulated spectra, consisting of a total of 387 intervals, with each interval corresponding to three consecutive spectral wavelengths. Subsequently, we constructed a label vector based on these intervals to characterize the distribution of BAL troughs within the spectrum. Specifically, if a BAL trough is present within an interval, the corresponding label value for that interval is marked as 1; otherwise, it is marked as 0 (as indicated by the red line at the top of Figure \ref{fig:ConstructSample} (c)).
In this way, we simplified the structure of the label data while retaining the key information about the BAL troughs in the spectrum, thereby providing more efficient data support for model training.

Regarding the division of the training and testing sets, we randomly selected 80\% of the samples from the final dataset, amounting to 160,000 spectra and their corresponding labels, to form the training set for the model. The remaining 20\% of the samples, which is 40,000 spectra and their labels, were then used for testing and evaluating the model.

\section{Methodology}\label{sec:deepLearning}

In this section, we will provide a detailed introduction to the \Med\ model used for detecting BAL troughs in quasar spectra. This model innovatively integrates a one-dimensional convolutional neural network (1D-CNN) and bidirectional long short-term memory networks (Bi-LSTM, \citealt{1997_Schuster_RNN_ITSP...45.2673S}), which enables effective extraction of both local and global features from spectral data. The 1D-CNN layer focuses on capturing local features in the spectrum, such as absorption trough profiles and continuum fluctuations, while the Bi-LSTM layers learn the global contextual dependencies of the spectrum through bidirectional sequence modeling (forward and backward propagation). This combined approach significantly improves the detection performance of BAL trough. In particular, when dealing with quasar spectra that have multiple BAL troughs, the model can more accurately identify and locate these features, thereby improving the accuracy and reliability of detection.

It is important to note that although 1D-CNN has been widely used in previous BAL automatic recognition studies \citep{2018_Busca_arXiv180809955B,2019_Guo_ApJ...879...72G,2025_Pang_AS}, the Bi-LSTM integration scheme proposed in this paper is the first sequence modeling-based work in this field. A key advantage of this approach lies in the Bi-LSTM’s ability to dynamically learn dependencies across arbitrary distances, a capability not inherent to CNNs. This characteristic is particularly crucial for detecting disjoint BAL troughs, which are common in our data. While a CNN with varying large kernels could be designed to capture long-range correlations, it requires careful manual tuning of kernel sizes and depths. In contrast, the Bi-LSTM adaptively learns the relevant context lengths from the data, leading to a more efficient and flexible architecture for modeling the complex, long-range dependencies present in BAL spectra. Furthermore, the Bi-LSTM’s capacity to simultaneously model both global long-term trends and local short-term fluctuations aligns closely with the comprehensive reasoning process astronomers employ in manual BAL identification. These combined strengths make it especially suitable for handling the multiple absorption features commonly found in quasar spectra. The following sections will first introduce the basic principles of LSTM/Bi-LSTM networks, and then provide a detailed description of the architecture and training strategy of \Med.

\subsection{LSTM Network}\label{ssec:LSTM}
LSTM network is a specialized type of recurrent neural network (RNN) initially proposed by \cite{1997_Hochreiter_LSTM}. It has been widely used for processing sequential data, including data analysis of spectral and time series \citep{2022_Hu_ApJ...930...70H,2023_Tabasi_ApJ...954..164T,2024_Luo_MNRAS.535.1844L}. Unlike traditional RNNs \citep{1997_Schuster_RNN_ITSP...45.2673S}, LSTM networks effectively addressed the vanishing and exploding gradient problems encountered when processing long sequences by incorporating memory cell states. Due to its strong memory capacity and ability to capture long-term dependencies, LSTM network is frequently employed in time-series forecasting and reliability prediction.

\begin{figure}[hpbt]
	\centering
	\includegraphics[width=\linewidth]{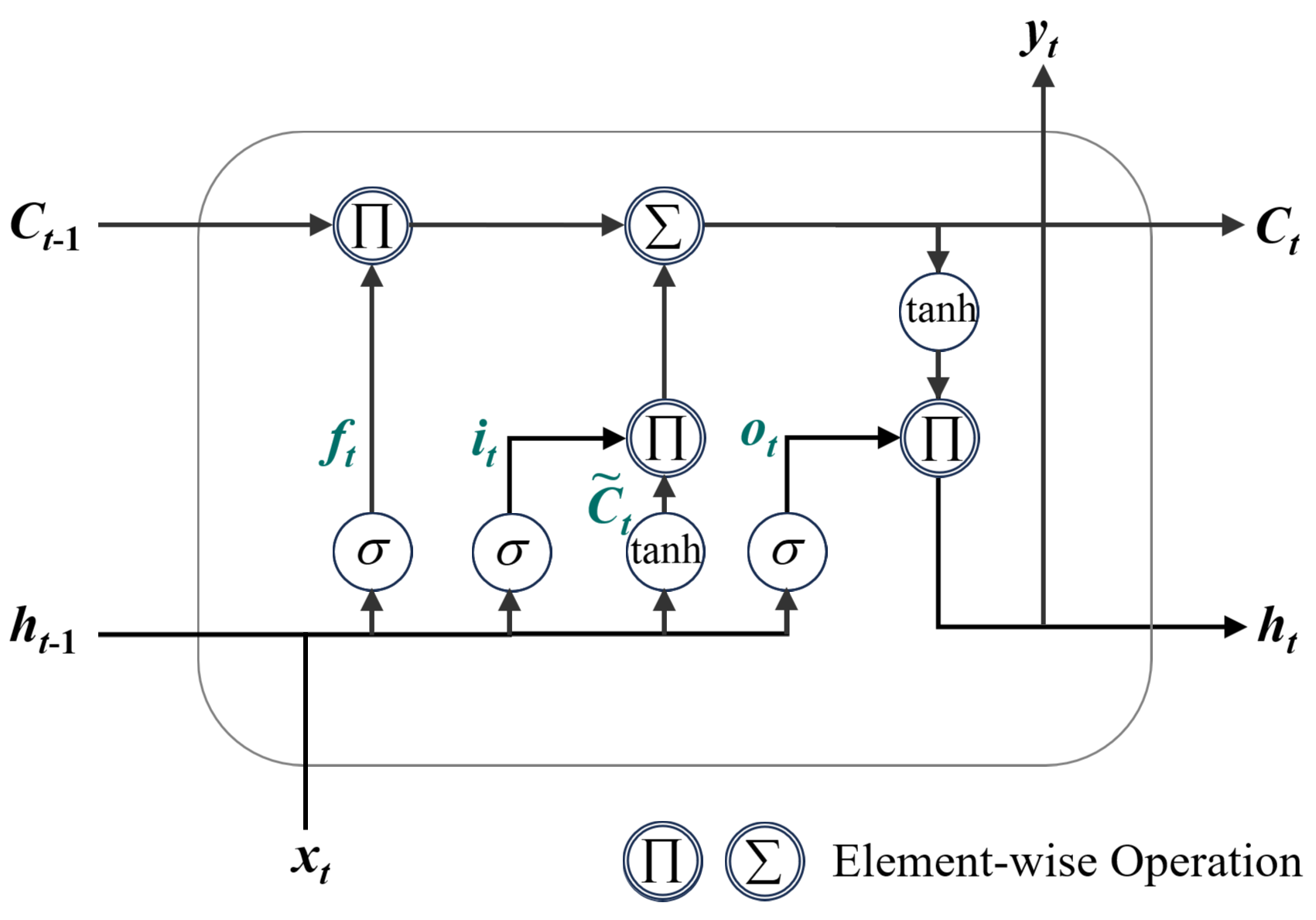}
	\caption{Schematic of the LSTM unit structure, illustrating the four interacting key components (input gate $i$, forget gate $f$, output gate $o$, and candidate memory cell state $\tilde{C}$) and their information flow. The mathematical expressions for the gates are provided in Equation (\ref{3}), while the temporal update mechanism of the cell state is strictly defined in Equation (\ref{7}). The arrows in the figure clearly indicate the direction of information flow, reflecting the dynamic gating logic of LSTM when processing sequential data.}
    
	\label{fig:LSTM}
\end{figure}

Similar to RNNs, LSTM networks employ a chain-like structure but feature a more sophisticated design in their repeating modules. By incorporating memory cells and gating mechanisms, LSTMs can effectively capture long-range dependencies. As illustrated in Figure \ref{fig:LSTM}, the control flow of a single LSTM timestep involves four core components: the input gate ($i$), forget gate ($f$), output gate ($o$), and candidate memory cell state ($\tilde{C}$). During timestep propagation, these elements operate in concert: the forget gate regulates which historical information to preserve, the input gate selects relevant features from the current input, the cell state synthesizes new and existing information to update memory, while the output gate governs the activation value for the current timestep. This integrated mechanism enables LSTMs to efficiently process sequential data and establish accurate input-output mappings.

\begin{figure*}[ht]
	\centering
	\includegraphics[width=\linewidth]{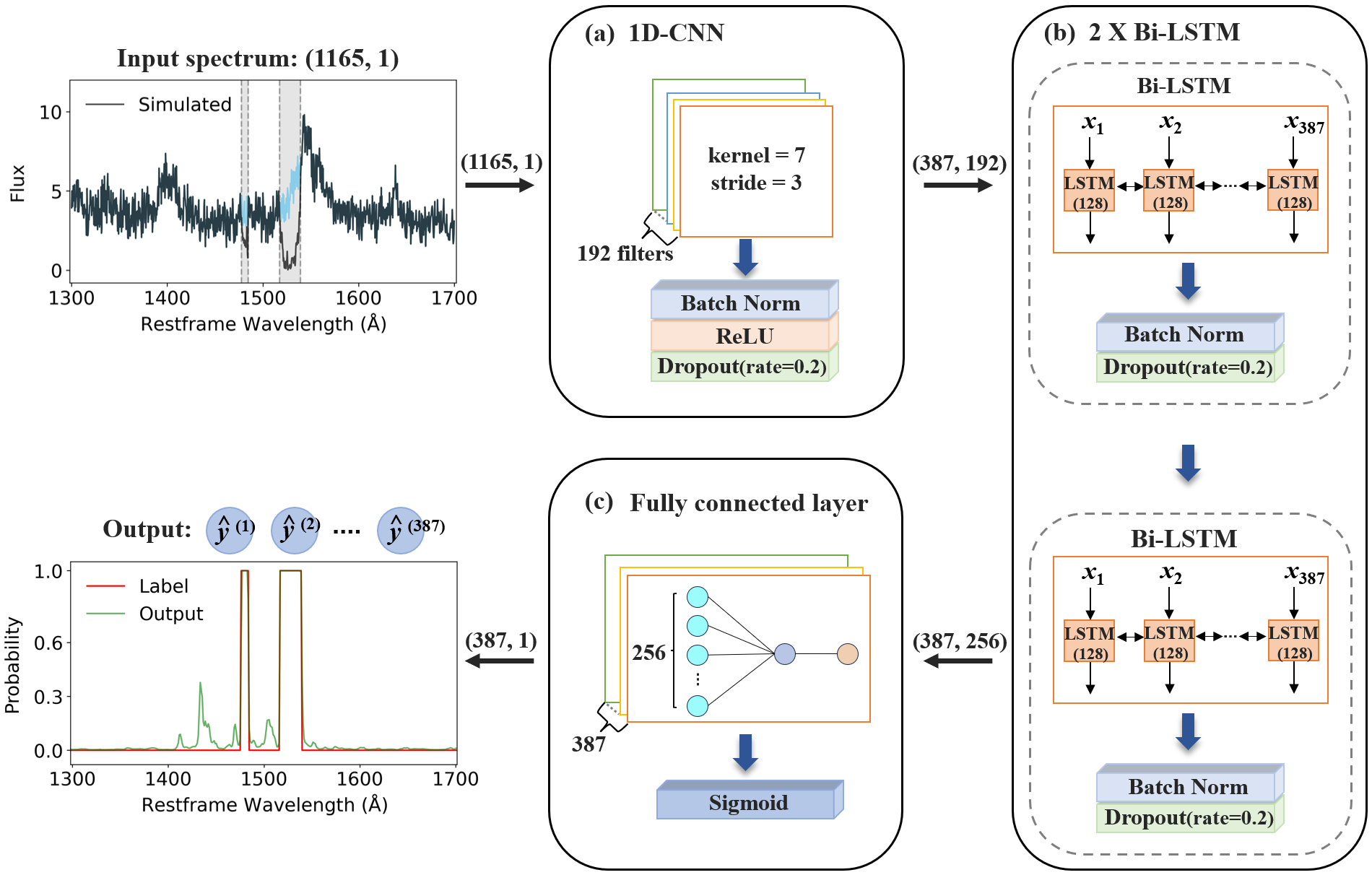}
	\caption{Architecture of the proposed \Med\ framework, comprising three core modules: (a) The 1D-CNN feature extractor processes the 1165-dimensional input spectrum through convolutional operations (kernel\_size=7, stride=3, 192 filters), transforming it into a 387×192 feature matrix, followed by batch normalization, ReLU activation, and dropout (rate=0.2) for feature refinement; (b) The Bi-LSTM module employs two bidirectional LSTM layers (128 hidden units each) to analyze temporal patterns, producing a 387×256 feature matrix. Each LSTM layer is followed by batch normalization and dropout for regularization; (c) The output module generates a 387-dimensional probability vector through a fully connected layer with sigmoid activation, where each element represents the presence probability of BAL troughs at the corresponding spectral position.}
   	\label{fig:StructureLSTM}
\end{figure*}

By applying the sigmoid ($\sigma$) and hyperbolic tangent (tanh) activation functions to the current input sequence $x_{t}$ and the previous hidden state $h_{t-1}$, the LSTM operations are updated at each time-step ($t$) according to the following equations:
\begin{equation}
    \begin{gathered}
        f_{t}=\sigma\left(W_{f} \cdot\left[h_{t-1}, x_{t}\right]+b_{f}\right), \\[0.7em] 
        i_{t}=\sigma\left(W_{i} \cdot\left[h_{t-1}, x_{t}\right]+b_{i}\right), \\[0.7em] 
        o_{t}=\sigma\left(W_{o} \cdot\left[h_{t-1}, x_{t}\right]+b_{o}\right), \\[0.7em] 
        \widetilde{C}_{t}=\tanh \left(W_{C} \cdot\left[h_{t-1}, x_{t}\right]+b_{C}\right), 
    \end{gathered}
    \label{3}
\end{equation}
where ($W_{f}, W_{i}, W_{o}, W_{C}$) and ($b_{f}, b_{i}, b_{o}, b_{C}$) are the weight matrices and bias weights, respectively. $\widetilde{C}_{t}$ indicates the candidate cell state. Then, utilizing the above equations to update the cell state $C_{t}$ and hidden state $h_{t}$ at the current time-step.
\begin{equation}
    \begin{gathered}
        C_{t} = f_{t} \odot C_{t-1} + i_{t} \odot \tilde{C}_{t}, \\[0.7em] 
        h_{t} = o_{t} \odot \tanh \left(C_{t}\right),
    \end{gathered}
    \label{7}
\end{equation}
where the $\odot$ symbol denotes an element-wise product. The initial values of $C_{0}$ and $h_{0}$ are both set to 0. This process illustrates how the LSTM regulates the input of new information and the flow of output information via its gating mechanism, thereby facilitating the effective maintenance of long-term memory and dependencies in sequential data. Through its structural design and gating mechanism, the LSTM enhances its performance and accuracy.

In addition, compared with most traditional machine learning methods, LSTM can automatically and effectively learn the dependencies between elements in the input sequence, thereby capturing the long-term dependencies and dynamic characteristics in sequential data. This capability gives LSTM a significant advantage in handling long sequence data.

The Bi-LSTM network is an enhanced variant of LSTM that incorporates two parallel LSTM layers: a forward layer processing the sequence chronologically and a backward layer processing it in reverse order. This bidirectional architecture enables simultaneous capture of both historical and future contextual features, thereby significantly improving the modeling of long-range dependencies in sequential data.

\subsection{\Med\ Architectures}\label{ssec:structure}

The primary objective of this study is to accurately identify BAL troughs in quasar spectra and determine their specific positions within the spectra, in order to obtain the associated velocity information. This task presents significant challenges because BAL troughs exhibit remarkable morphological and positional diversity across different spectra, and are often obscured by interference from other spectral features (such as emission lines and continuum spectra). To address these challenges, we developed a specialized neural network model named \Med\, which innovatively combines the strengths of 1D-CNN and Bi-LSTM with 1D-CNN extracting local spectral features while Bi-LSTM effectively captures long-term dependencies in spectral sequences. The neural network architecture was implemented using the Keras \footnote{\url{https://keras.io/about/}} \citep{2015_chollet_keras} and TensorFlow \citep{abadi2016tensorflow} frameworks, with Keras serving as the high-level API for TensorFlow. Furthermore, to ensure reproducibility and enable community access, both the source code and the generated catalogs have been made publicly available on our GitHub repository \footnote{\url{https://github.com/zjluo-code/BALNet}}.

The network architecture of \Med, illustrated in Figure \ref{fig:StructureLSTM}, follows an end-to-end deep learning approach with three core modules connected in series. First, the 1D-CNN feature extraction module processes the input 1165-dimensional spectral vector. Using a single 1D convolutional layer (kernel\_size = 7, stride = 3, 192 filters), it transforms the input into a 387×192 feature matrix, where 387 corresponds to the sequence length and 192 denotes the number of feature channels. This module includes a complete feature optimization process: a Batch Normalization layer (Batch Norm) to stabilize network training, a ReLU activation function to introduce non-linearity \citep{2015_Xu_arXiv150500853X}, and a Dropout layer (rate = 0.2) to prevent overfitting \citep{2012_Hinton_arXiv1207.0580H,2014_Zaremba_arXiv1409.2329Z}.

Next, the Bi-LSTM module receives the 387×192 feature matrix and processes it through two layers of bidirectional LSTM. Each LSTM layer contains 128 hidden units and the bidirectional outputs are concatenated to form a 256-dimensional vector, resulting in the final output of a 387×256 feature matrix. This module also applies Batch Norm and Dropout (rate = 0.2) for regularization.

Finally, the output module transforms the 387×256 feature matrix into a 387-dimensional probability output vector through a fully connected layer and a sigmoid activation function. Each element in the vector represents the probability of the presence of a BAL trough at the corresponding position. Typically, a threshold of 0.5 is used for binary classification to determine the existence of a BAL trough point. In addition, in the model, we regard the continuous occurrence of five or more BAL trough points as a confirmation signal for a BAL trough. 

The entire network performs end-to-end processing from raw spectral inputs to both classification and positional (velocity) parameter prediction, enabled by strict dimensional control and consistent hyperparameter settings throughout the architecture.

The loss function of the \Med\ model employs binary cross-entropy loss (BCE) \footnote{\url{https://www.tensorflow.org/api_docs/python/tf/keras/losses/BinaryCrossentropy}} . This loss function optimizes the model's performance by calculating the difference between the model's predicted output and the true labels.

After model construction, we trained the network using 160,000 normalized spectral entries from the training set. Each spectrum was normalized by its maximum value prior to input. All experiments were performed on an NVIDIA RTX 3090 GPU. We employed the Adam optimizer \citep{kingma2014adam} with $\beta_1 = 0.5$, $\beta_2 = 0.999$, a learning rate of 0.0001, and a batch size of 256. The training completed in 200 epochs, with each epoch requiring $\sim$ 42 seconds, resulting in a total training time of approximately 2.4 hours.


\section{{Model Performance Evaluation}}\label{sec:resultanddiscussion}

In this section, we evaluate how effectively our trained \Med\ model performs when detecting BAL troughs and measuring their locations in simulated quasar spectra using the test dataset.

\subsection{Evaluation Metrics}\label{ssec:metrics}

We quantitatively evaluate \Med's BAL trough detection performance using three standard metrics: completeness (recall), purity (precision) and F1-score. These metrics constitute a robust evaluation framework where: (1) completeness measures the model's ability to identify all relevant troughs, (2) purity quantifies the correctness of detected troughs, and (3) the F1-score balances these complementary aspects. The metrics are formally defined as:
\begin{equation}
    \begin{aligned}
        & \text{Completeness} = \frac{\rm{TP}}{\rm{TP} + \rm{FN}},\\[0.7em] 
        & \text{Purity} = \frac{\rm{TP}}{\rm{TP}+ \rm{FP}},\\[0.7em] 
        & \text{F1-score} = 2 \times \frac{\mathrm{Purity} \times \mathrm{Completeness}}{\mathrm{Purity} + \mathrm{Completeness}},
    \end{aligned}
    \label{combined_lstm}
\end{equation}
where, TP (true positive) represents the number of BAL troughs that are correctly detected by the model, FP (false positive) represents the number of non-BAL troughs that are incorrectly detected as BAL troughs by the model, and FN (false negative) represents the number of actual BAL troughs that are not detected by the model.

Furthermore, to fully evaluate the performance of \Med\ in different decision thresholds, we employ the area under the precision-recall curve (\mbox{AU-PRC}) as a complementary evaluation metric. AU-PRC is calculated by integrating the area bounded by the precision-recall (PR) curve and the axes, providing a comprehensive measure of the model's performance stability under varying threshold configurations. This metric ranges from $\rm [0, 1]$, where 1 represents ideal performance and 0.5 corresponds to random guessing. A higher AU-PRC value indicates that the model maintains high precision while simultaneously achieving high recall, confirming the superior overall detection capability.

To evaluate the performance of our model in estimating BAL velocity parameters based on location measurements, we used the label values of the simulated data as a benchmark and systematically compared the velocity parameters predicted by the model \Med\ with these true values. To comprehensively quantify the measurement quality, we employed three metrics: the fraction of catastrophic outliers ($f_{out}$), the normalized median absolute deviation ($\rm \sigma_{NMAD}$), and the bias in the BAL velocity parameter measurements ($\mathrm{bias}$) to assess the quality of the model's velocity parameter measurements \citep{2008Brammer_ApJ...686.1503B,2024_Luo_MNRAS.535.1844L,2024_Luo_MNRAS_531}.

\begin{table*}[hpt]
\caption{Evaluation metrics of BAL trough detection by \Med, using the optimal \PROB\ threshold of 0.3.}               \label{table:metrics}      
\centering                         
	\begin{tabular}{c c c c c}        
	\toprule 
	 &  Completeness (\%) & Purity (\%) &  F1-score (\%) & AU-PRC \\ 
	\hline                        
	Train  & 83.1 & 90.5 & 86.6 & 0.92 \\ 
	Test   & 83.0 & 90.7 & 86.7 & 0.92 \\             
	\hline   	\hline                                 
	\end{tabular}
\end{table*}

Among these metrics, $f_{out}$ measures the proportion of predictions that fall outside an acceptable error range, indicating potential significant deviations from true values. A velocity parameter estimate is considered a catastrophic outlier if it meets the following condition:
\begin{equation}
	\frac{|\Delta v|}{1+\left|v_{\rm true}\right|}>0.15,
\end{equation}
 where 
\begin{equation}
	\Delta v =v_{\rm pred}-v_{\rm true},
\end{equation}
$v_{\rm true}$ is the reference velocity used as the "ground truth" benchmark, and $v_{\rm pred}$ is the velocity parameter predicted by the model, both in units of 1000 $\rm{km/s}$. $\rm \sigma_{NMAD}$ quantifies the dispersion of predicted velocity values relative to ground truth, serving as  a robust measure of estimation precision. It is defined as:

\begin{equation}
	{\rm \sigma_{NMAD}}=1.48\times {\rm median}\left(\left|\frac{\Delta v - {\rm median}(\Delta v)}{1+\left|v_{\rm true}\right|}\right|\right).
\end{equation}
The $\mathrm{bias}$ assesses whether there is a systematic tendency for the model to overestimate or underestimate the velocity parameters relative to the truth, and it is typically calculated using the following formula:
\begin{equation}
	{\rm bias} = {\rm median}\left(\frac{\Delta v}{1+\left|v_{\rm true}\right|}\right).
\end{equation}

\begin{figure}[ht]
	\centering
	\includegraphics[width=0.75\linewidth]{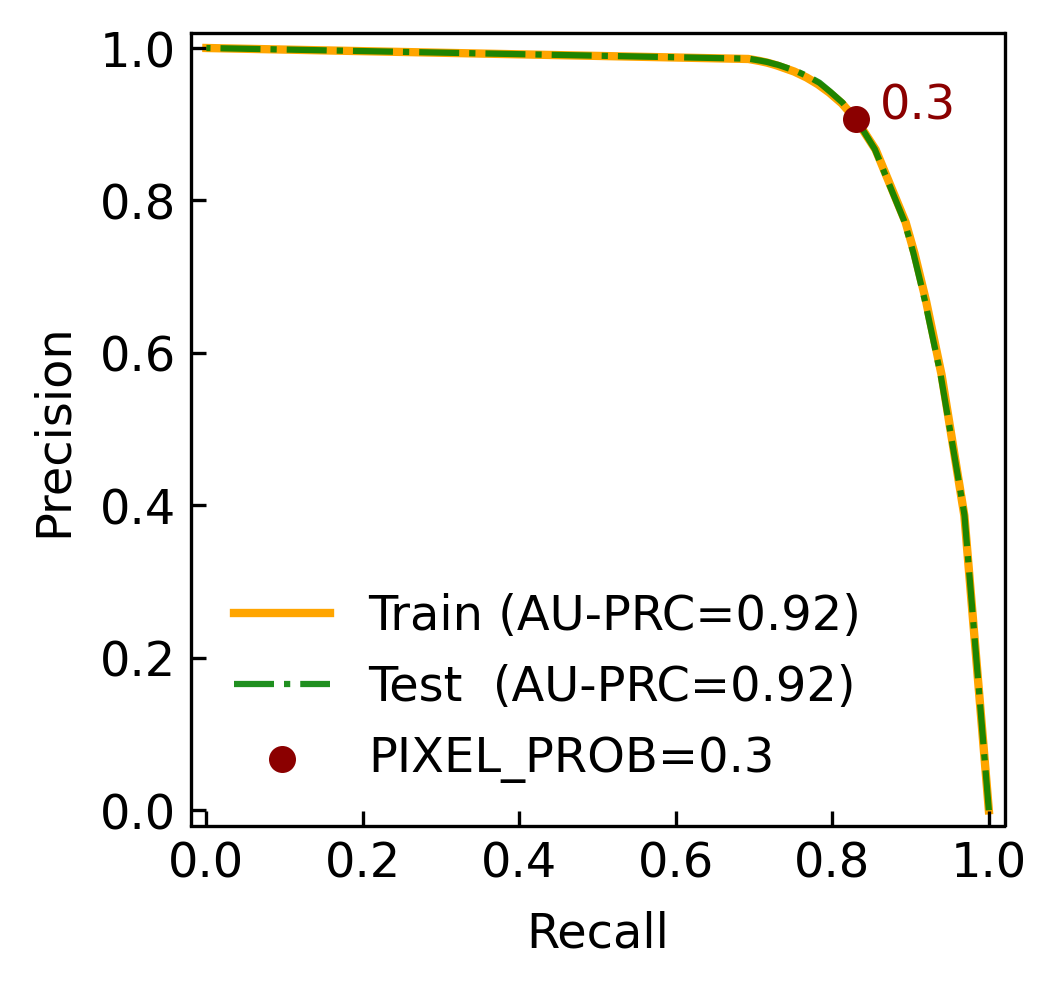}
	\caption{ Precision-recall (PR) curves for the training set (orange) and the testing set (green). The brown point marks the probability threshold at which the model achieves its optimal performance (Best F1-score, \PROB =0.3). The AU-PRC serves as a quantitative measure of the model's performance.}
	\label{fig:ACC_num}
\end{figure}

\begin{figure*}[t]
	\centering
	\includegraphics[width=\linewidth]{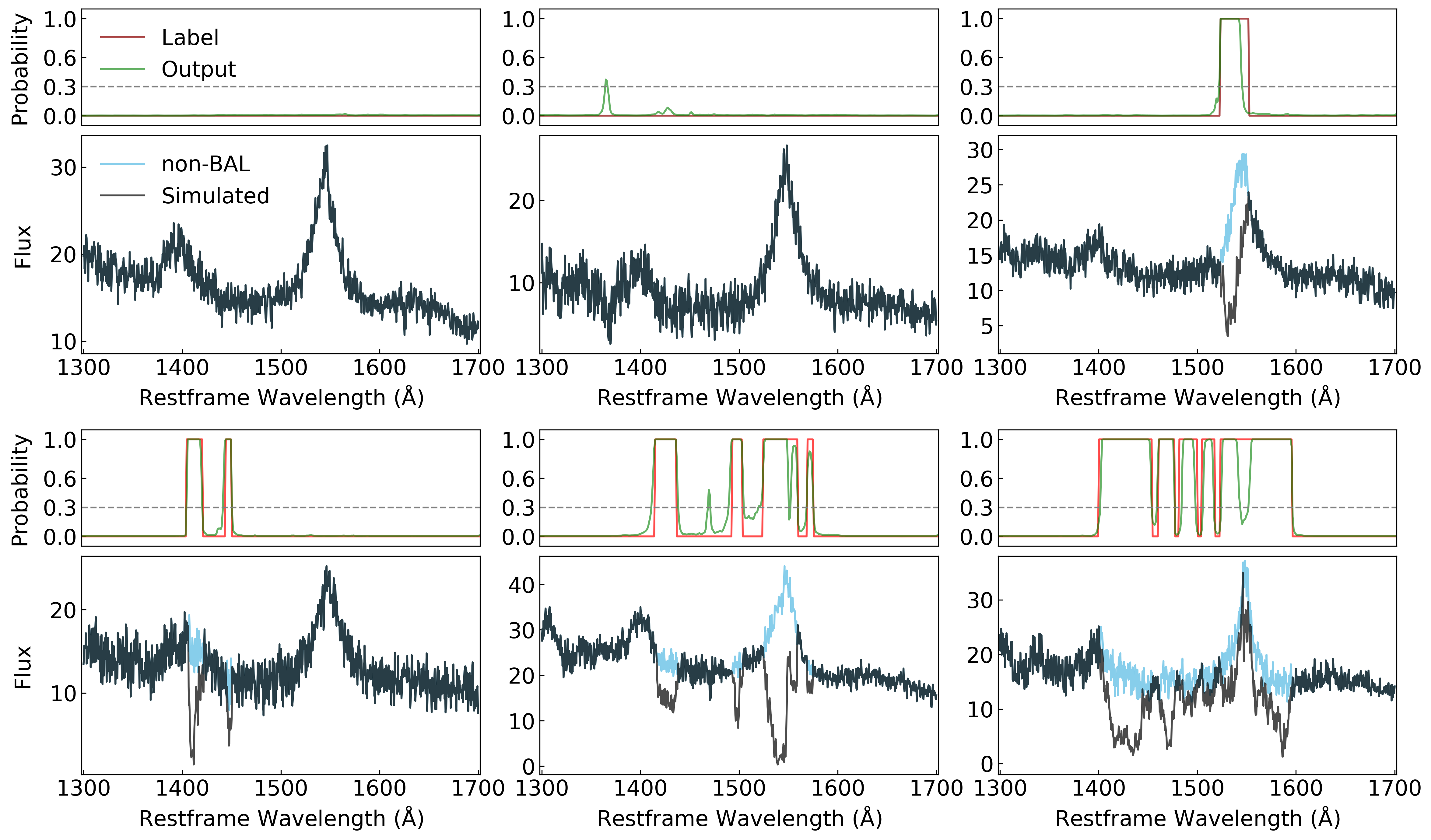}
	\caption{Examples from the test set and corresponding model predictions. 
	In each panel, the top row displays the ground-truth label vectors in red and the model-predicted probabilities in green, scaled between 0 and 1. 
	The bottom row shows the simulated spectra in black and their corresponding \NONBAL\ background spectra in blue; 
	the difference between the two reveals the BAL troughs embedded within the simulated spectra.}
	\label{fig:samples}
\end{figure*}

\subsection{BAL Trough Detection} \label{ssec: detection}
As can be seen from the architecture of the \Med\ (Figure \ref{fig:StructureLSTM}), for each quasar spectrum, the model outputs a 387-dimensional probability vector. Therefore, we can identify BAL troughs in quasar spectra by setting an appropriate threshold (PIXEL\_PROB). Although a higher threshold can improve the purity of identification, this is often at the expense of completeness. To obtain the optimal value of PIXEL\_PROB, we analyzed the PR curves of the model on the training and test sets and calculated the AU-PRC values. As shown in Figure \ref{fig:ACC_num}, the AU-PRC values reached 0.92. This near-ideal performance (where 1.00 indicates perfect model performance), along with the close alignment between the training and testing PR curves, confirms that our model has good generalization ability and has not overfitted. Considering the balance between precision and recall, we ultimately determined the optimal pixel probability threshold to be 0.3 based on the F1-score.

At this threshold, our trained model achieved a BAL trough identification accuracy of 90.7\% on the testing set. Table \ref{table:metrics} summarizes the detailed performance metrics of the model on both the training and testing sets, including completeness, purity, and F1-score. As shown in the table, the \Med\ model performs consistently well across both datasets, achieving a completeness of around 83\%, a purity of approximately 91\%, and an F1-score of about 87\%. These results demonstrate the model’s robustness and effectiveness in detecting BAL troughs in quasar spectra. Figure \ref{fig:samples} presents representative examples in which the trained model successfully identifies spectra containing zero, one, or multiple BAL troughs.

The presence of BAL troughs in quasar spectra serves as a key diagnostic for identifying \BALQSOs. Accordingly, our model can be applied to spectral data to classify sources as BAL or \NONBALs. Evaluation on the testing set shows that the model achieves a completeness of 90.8\% and a purity of 94.4\% in \BALQSO\ classification. These results indicate that the model effectively detects the majority of true \BALQSOs\ while maintaining a very low false-positive rate, demonstrating its robustness and reliability in distinguishing \BALQSOs\ from general quasar populations.

\begin{figure*}[hptb]
	\centering
	\includegraphics[width=\linewidth]{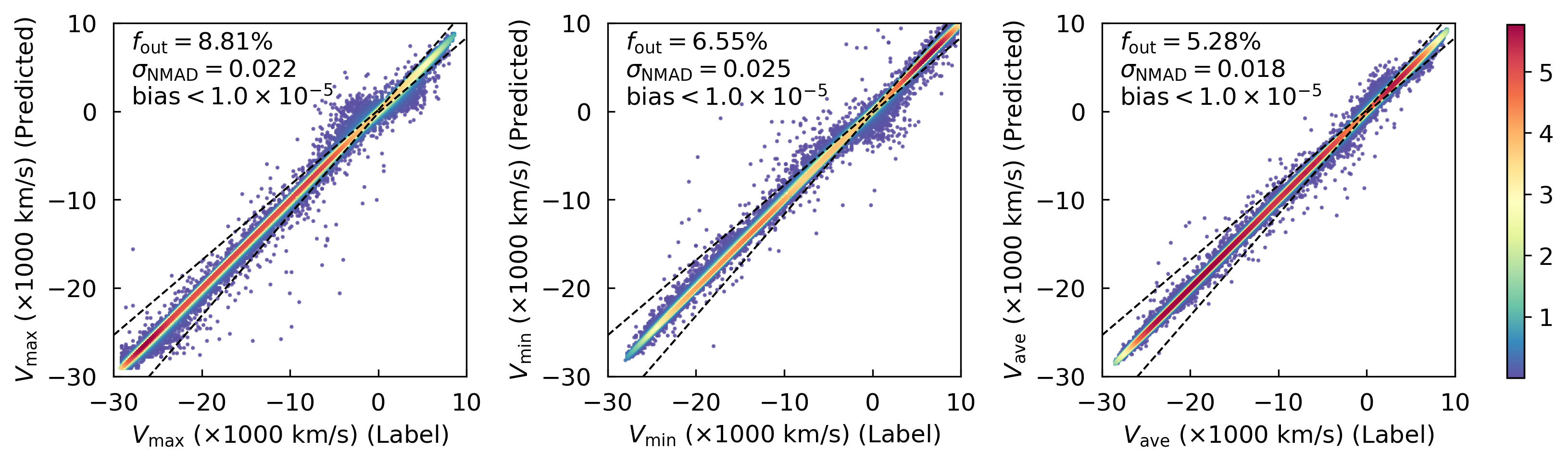}
	\caption{Comparison between the ``ground-truth" labels and the predicted BAL velocity parameters from simulated spectra.
	Model performance is assessed using three metrics: the fraction of catastrophic outliers ($f_{\mathrm{out}}$), the normalized median absolute deviation ($\sigma_{\mathrm{NMAD}}$), and the systematic velocity bias ($\mathrm{bias}$).
	The solid line indicates perfect agreement between predicted and label values, while the dashed lines mark the $\pm$0.15 threshold set by the spectral resolution; data points lying outside this range are considered catastrophic outliers.
	The color scale reflects the density distribution of data points.}	\label{fig:density}
\end{figure*}

\begin{table*}[hpt]
\caption{Comparing the completeness and purity of \Med\ with previous works in \BALQSO\ classification. Note that different studies employ distinct datasets and labeling strategies.}  \label{table:comparision}     
\centering                                   
\begin{tabular}{c c c c c c}      
	\toprule   
	Model     &  Completeness (\%)   &  Purity (\%) & Train: Test & Data & Reference \\
	\hline
    \Med\ (CNN + Bi-LSTM) & 90.8 & 94.4 & 8:2 & Mock data & this work \\
	\texttt{QuasarNet} (CNN) & 98.0 & 77.0 & 8:2  & DR12 &  \cite{2018_Busca_arXiv180809955B} \\
	PCA-echenced CNN & 97.4 &  40.0 & 9:1  & DR12 &  \cite{2019_Guo_ApJ...879...72G}\\
	\texttt{FNet II} (ResNet + CNN) & 99.0 & \_ & 9:1  & DR16/DR17 & \cite{2024_Moradi_MNRAS.533.1976M} \\   
	PCA + XGBoost & 97.7 & 96.2 & 9:1  & DR16 & \cite{2024_Kao_PASJ...76..653K} \\
	\hline    \hline
\end{tabular}
\end{table*}

\subsection{BAL Trough Velocity Measurements}\label{ssec: measure}

Based on the output vector of the \Med\ model, each element corresponds to a specific position in the input spectrum. This positional correspondence enables the determination of both the location and the width of BAL troughs in quasar spectra, from which three key velocity parameters are extracted: \( V_{\mathrm{max}} \) and \( V_{\mathrm{min}} \), and \( V_{\mathrm{ave}} \). Specifically, \( V_{\mathrm{max}} \) and \( V_{\mathrm{min}} \) denote the maximum (blueward) and minimum (redward) velocities of the BAL trough, respectively, while \( V_{\mathrm{ave}} \) represents its average velocity.
These parameters quantify the kinematic properties of BAL outflows and serve as important diagnostics of quasar physical conditions, such as the dynamics and geometry of the outflowing material. 

Figure \ref{fig:density} presents  the comparison  between the predicted and "ground-truth" label values of the velocity parameters on the testing set. 
As shown, the predicted values closely track the overall trend of the label values, indicating that the model effectively captures the kinematic properties of BAL troughs.
Quantitatively, the fraction of catastrophic outliers ($f_{\mathrm{out}}$) dose not exceed approximately 9.0\%, suggesting that while a small number of outliers exist, the model maintains high predictive accuracy overall. 
Furthermore, the normalized median absolute deviation ($\rm \sigma_{NMAD}$) is below 0.03, and the systematic bias in the measurement of the velocity ($\rm bias$) is less than 10$^{-5}$, further validating the robustness and high reliability of the model in the predicting of the velocity parameters.

\subsection{Compare With Other Works}

In previous studies (see Table \ref{table:comparision}), \texttt{QuasarNet} \citep{2018_Busca_arXiv180809955B} and \texttt{FNet II} \citep{2024_Moradi_MNRAS.533.1976M} have been proposed as four-class classifiers designed to distinguish stars, galaxies, quasars, and \BALQSOs. These models fundamentally rely on the BAL flags provided in the SDSS catalog to label training samples and evaluate performance by comparing predictions against these flags. However, known inaccuracies in the BAL flags in SDSS, combined with significant class imbalance—\BALQSOs\ being far less numerous than other object types—may compromise the validity of the resulting performance metrics. 
For these reasons, although \texttt{QuasarNet} achieves a high completeness of 98.0\%, its relatively low purity of 77.0\% reflects a significant false positive rate, likely caused by the scarcity of BAL training examples.
Meanwhile, \cite{2024_Kao_PASJ...76..653K} found that among various dimensionality-reduction methods and machine-learning classifiers, the combination of PCA and the XGBoost classifier represents the pinnacle of efficacy in the \BALQSO\ classification task, boasting impressive accuracy rates of 97.60\% by 10-fold cross-validation and 96.92\% on the external testing set.
To improve label accuracy, \cite{2019_Guo_ApJ...879...72G} refined their labels through multiple iterations involving training, prediction, and visual re-inspection of ambiguous cases. Their PCA-based CNN method achieves completeness comparable to that of \cite{2018_Busca_arXiv180809955B}, while effectively capturing narrower absorption troughs with widths less than 2000~ km/s, as well as troughs extending to the center of the \ion{C}{4} emission line.

\begin{figure*}[hptb]
	\centering
	\includegraphics[width=\linewidth]{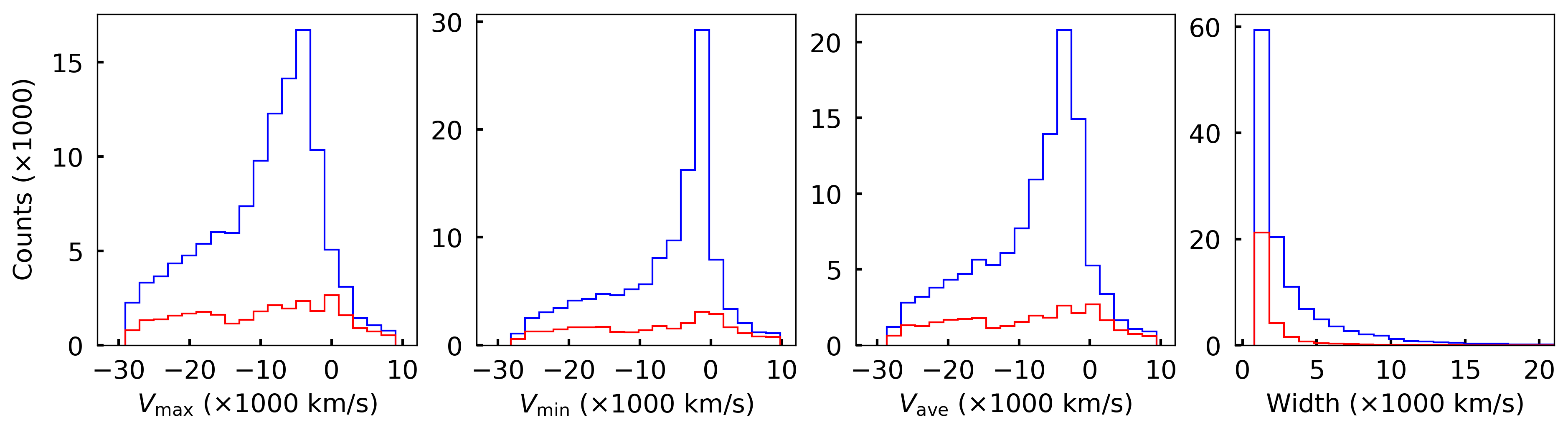}
	\caption{Velocity and width distributions of all BAL troughs (blue) identified by \Med\ in the \DR\ spectra.
	The red curves highlight the BAL troughs corresponding to newly identified \BALQSOs\ that are absent from the \DR\ catalog.
	}\label{fig:BALVecli}
\end{figure*}

In contrast, \Med\ exhibits a more balanced performance,  achieving a completeness of 90.8\% and a purity of 94.4\%, which demonstrates its robustness in distinguishing \BALQSOs\ from \NONBALs\ in mock data.
The remarkable performance of \Med\ can be attributed to its innovative architecture, which cleverly integrates 1D-CNN and Bi-LSTM networks. The 1D-CNN specializes in extracting local spectral features such as absorption troughs and continuum variations, whereas the Bi-LSTM effectively captures global spectral patterns by modeling long-range dependencies. This dual feature extraction strategy not only enhances feature robustness but also significantly improves the accuracy of BAL trough detection. More importantly, unlike general-purpose classifiers like \texttt{QuasarNet} and \texttt{FNet II}, \Med\ is explicitly designed for BAL trough detection—a specialized design that is crucial to its superior purity.
Nevertheless, it is important to emphasize that the completeness and purity metrics reported in Table \ref{table:comparision}  reflect only the model’s performance on the training dataset, and the quality of the labels in the training data directly determines the model’s generalizability to real observational data. 
In this work, the BAL troughs and unabsorbed wavelength regions in the datasets used for \Med\ training and testing are precisely labeled.

\section{Application to DR16Q}\label{sec: resultDr16}

As demonstrated in the evaluation using the mock dataset, \Med\ exhibits excellent  performance in both BAL trough detection and velocity parameter estimation. To further assess its practical effectiveness, we applied the trained model to the quasar spectral data released by the \DR.
The trained \Med\ model (with the optimal probability threshold \PROB\ = 0.3) was applied to a total of 446,839 quasar spectra from the \DR\ dataset within  the redshift range of $1.5 \le$ z $\le 5.7$.
All spectra were preprocessed  by performing linear interpolation to 1165 uniformly spaced sampling points within the wavelength range [1300, 1700]~\AA, followed by three-point moving average smoothing, to ensure consistency with the format of the simulated training set.

A total of 117,626 BAL troughs have been identified by \Med\ in 91,164 quasars, 
indicating that 20.4\% of the sources in the \DR\ sample are classified as \BALQSOs.
The velocity and width distributions of these BAL troughs are shown in Figure~\ref{fig:BALVecli}.
As shown, our search covers a velocity range from blueshifts of 29,000~ km/s to redshifts of 10,000~ km/s,
which is significantly broader than the velocity ranges typically explored in previous studies,
where the focus was generally on blueshifts between 25,000~ km/s and 3,000~ km/s or up to 0~ km/s.
As a result, our sample includes a more diverse set of BAL troughs, comprising both high-velocity blueshifted and redshifted BAL troughs.
For example, we identify 3,379 high-velocity blueshifted BAL troughs with mean velocities ($V_{\rm ave}$) exceeding 25,000~ km/s,
as well as 10,370 redshifted BAL troughs with $V_{\rm ave} \ge 0$~ km/s.

According to the \DR\ catalog, \citet{2020_Lyke_ApJS..250....8L} identified a total of 99{,}856 \BALQSOs\ with $\mathrm{BAL\_PROB} \geq 0.75$. These objects can be categorized into two distinct groups based on their \ion{C}{4} BAL properties: (1) 23{,}994 quasars with $\mathrm{BI\_{CIV}} > 0$, and (2) 75{,}702 quasars with $\mathrm{BI\_{CIV}} = 0$ and $\mathrm{AI\_{CIV}} > 0$. 
The former group serves as the parent sample from which we extract \ion{C}{4} BAL troughs, as described in Section~\ref{ssec:DataPreparation}. Each trough in the spectra of these sources is precisely measured using the pair-matching method.
Among these \BALQSOs, \Med\ successfully recovered 98.3\% of the sources. Of the 412 missed cases, the pair-matching method determined that 153 objects do not exhibit any BAL features, implying that the true missing  rate of \Med\ is only 1.1\%.
For the latter, comparison with the \Med\ catalog reveals that 33,243 sources —  accounting for 43.9\% of the total — were rejected. 
According to the definitions of $\mathrm{BI\_{CIV}}$ and $\mathrm{AI\_{CIV}}$ in \citet{2020_Lyke_ApJS..250....8L}, these sources exhibit absorption troughs confined to the blueshifted velocity range of 0 - 3{,}000~ km/s. 
This suggests that \Med\ may still have limitations in detecting BAL troughs at low velocities, particularly  those overlapping the \ion{C}{4} emission line. In addition, \Med\ also newly identified 25,123 \BALQSOs. The velocity and width distributions of their BAL troughs are shown as red curves in Figure~\ref{fig:BALVecli}. As illustrated, the newly detected BAL troughs are relatively uniformly distributed in velocity space, while the majority (77.8\%) have widths ($V_{\rm min}-V_{\rm max}$) below 2,000 ~km/s.

\begin{figure}[ht]
	\centering
	\includegraphics[width=0.86\linewidth]{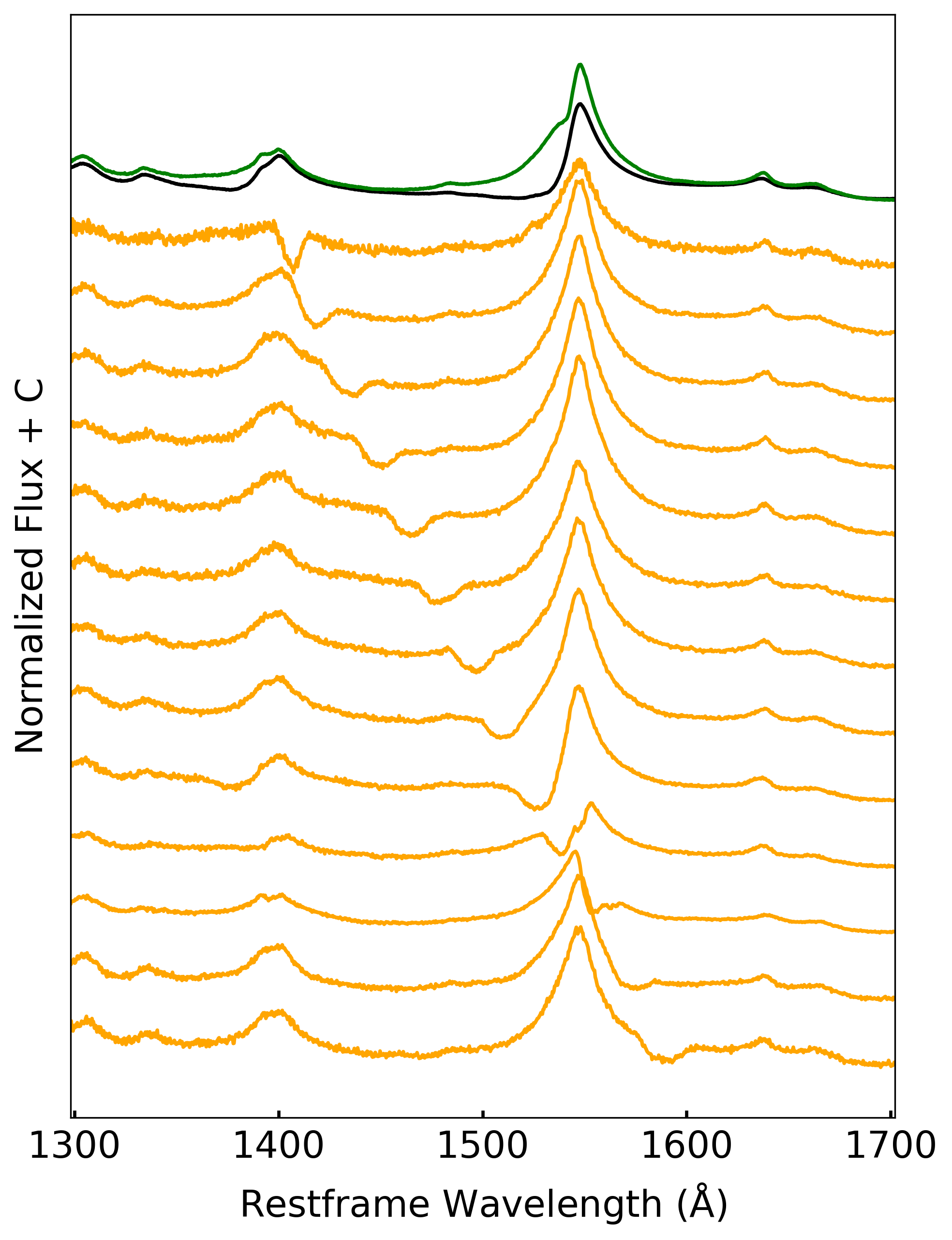}
	\caption{Composite spectra of different \BALQSO\ subsamples.
		The black and green curves represent the \BALQSOs\ in \DR\ that are recovered and rejected by \Med, respectively.
		The orange curves show a sequence of composite spectra for the \BALQSOs\ newly identified by \Med.}
	\label{fig:Composite}
\end{figure}

\begin{figure*}[ht]
	\centering
	\includegraphics[width=0.78\linewidth]{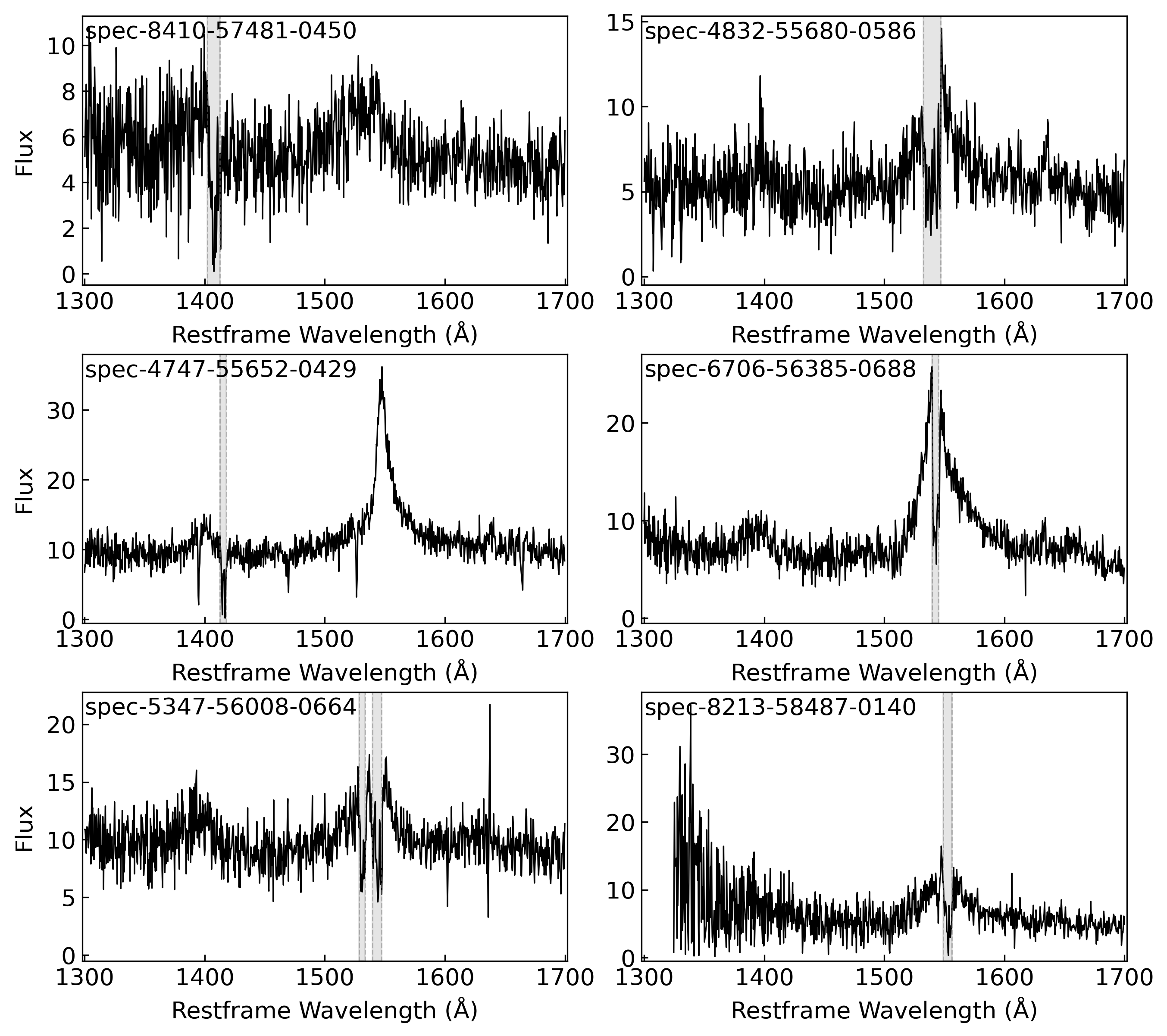}
	\caption{Examples of newly identified \BALQSOs. Each panel displays the observed spectrum in black, with the BAL troughs detected by \Med\ highlighted in gray-shaded regions; the last panel features a redshifted absorption trough.}
	\label{fig:NewSample}
\end{figure*}

To further assess the reliability of \BALQSO\ identification, Figure~\ref{fig:Composite} compares the composite spectra of the aforementioned \BALQSO\ subsamples. The black curve represents the composite spectrum of 66,041 \BALQSOs\ identified by both \Med\ and \DR, displaying prominent BAL troughs. 
In contrast, the green curve — corresponding to 33,655 \BALQSOs\ classified by \DR\ but rejected by \Med\ — shows only an extremely weak absorption feature on the blue wing of the \ion{C}{4} emission line.
Given the relatively uniform velocity distribution of the newly detected BAL troughs by \Med, we construct a series of composite spectra for thses \BALQSOs\ with different average velocities ($V_{\mathrm{ave}}$), using velocity bins of 3,000~ km/s. These are shown in orange.
As shown, BAL troughs appear sequentially in the composite spectra from high to low velocities, transitioning from blueshifted to redshifted regions. This clearly confirms that the newly detected features are real and significantly more prominent than those in the \Med-rejected sample, affirming their nature as genuine BAL troughs in the observed spectra. Figure \ref{fig:NewSample} shows spectra randomly selected from these newly identified \BALQSOs, which exhibit clearly significant BAL troughs.

\begin{figure*}[ht]
	\centering
	\includegraphics[width=0.78\linewidth]{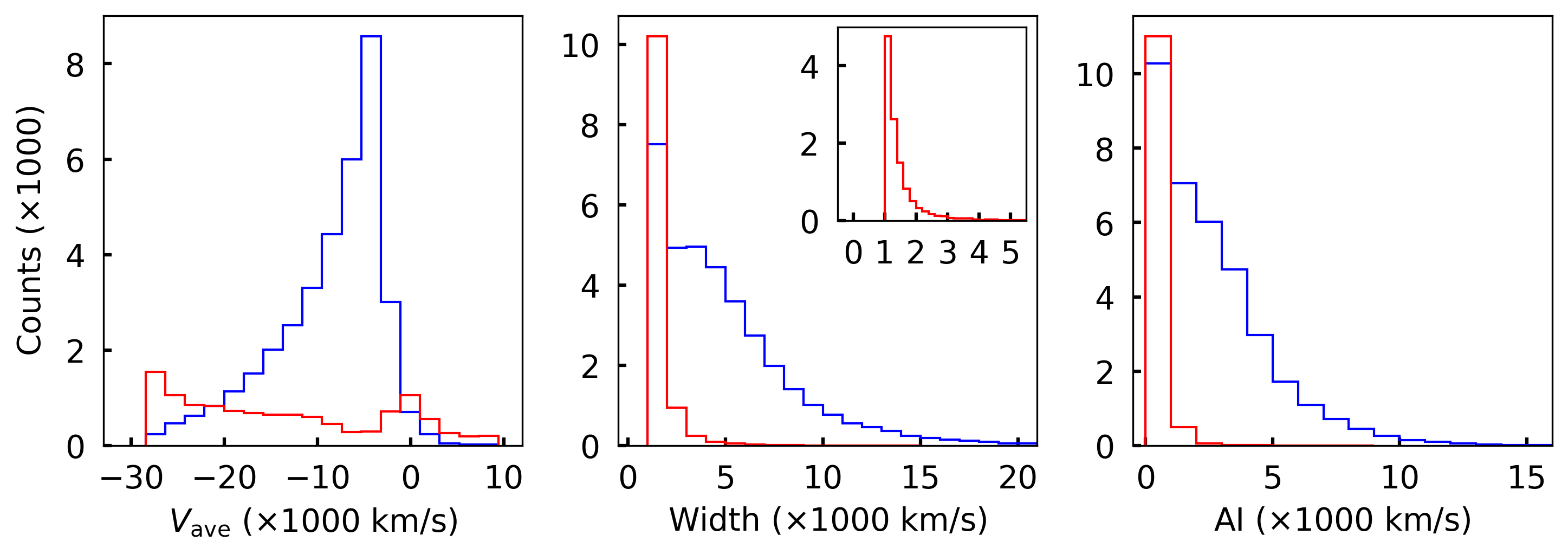}
	\caption{Comparison between BAL troughs recovered (blue) and those missed (red) by \Med\ in the BAL pattern library. In the middle panel, the upper-right inset presents the results obtained from a finer binning (200 km/s) of the missed BAL troughs.}
	\label{fig:bias}
\end{figure*}

Although the composite spectra have revealed that the rejected \BALQSOs\ display only weak absorption features on the blue wing of the \ion{C}{4} emission line, while the newly identified sources exhibit prominent BAL troughs — demonstrating the robustness and reliability of the \Med\ method — we still aim to further evaluate potential selection biases inherent in this approach.
As described in Section 2.1, we identified a total of 47,267 BAL troughs and constructed the BAL pattern library. In Figure~\ref{fig:bias}, we compare the velocity and width distributions of the troughs recovered by \Med\ with those that were missed. It is clear that the missed troughs are not uniformly distributed in velocity space. A disproportionately large fraction of them occur within the velocity range dominated by the \ion{C}{4} emission line and near the extreme high-velocity end close to the measurement limits. This suggests that, when generating mock datasets, it may be necessary to increase the number of targets in these specific velocity regions to ensure sufficient sampling. Moreover, the missed troughs generally exhibit weaker absorption strengths (AI) and narrower widths, with approximately 90\% having widths below 2,000~km/s, and nearly half of them narrower than 1,200~km/s.
In addition, the spectra of these missed absorption troughs generally have lower spectral quality, with 55\% of them exhibiting a signal-to-noise ratio below 3.0. During detection processing, we applied a three-point moving average smoothing to the original spectra, and to reduce computational costs, the size of the probability vector labels was also compressed when generating the training data. 
These compromises reduced the effective velocity resolution, causing some narrow troughs to fall below the detection threshold and be discarded. Additionally, the spectral signal-to-noise ratio also partially affects the detection results.


\section{Summary}\label{sec:conclusions}
In this study, we develop \Med, an innovative deep learning framework that integrates a 1D-CNN and Bi-LSTM networks to automatically detect \ion{C}{4} BAL troughs in quasar spectra. Unlike conventional approaches that are typically limited to \BALQSO\ classification, \Med\ not only identifies \BALQSOs\ but also directly measures the velocities of their BAL troughs. 
To ensure robust training and evaluation, we constructed a more comprehensive simulated dataset by combining \NONBAL\ spectra and BAL troughs, both meticulously derived from the \DR\ data.
Quantitative evaluations demonstrate the excellent performance of \Med\ on the testing set, achieving a completeness of 83.0\% and a purity of 90.7\% in BAL trough detection (F1-score = 86.7\%), as well as 90.8\% completeness and 94.4\% purity in \BALQSO\ identification. The velocity measurements of the detected BAL troughs exhibit strong consistency with the ground-truth labels, confirming both the accuracy and reliability of \Med.

Applied to 446,839 quasar spectra within the redshift range of 1.5 $\le$ z $\le$ 5.7 from the \DR, \Med\ identifies 91,164 \BALQSOs, accounting for 20.4\% of the sample and providing more comprehensive BAL trough coverage (8.8\% are redshifted BAL systems). 
Compared to the DR16Q-classified \BALQSOs\ (BAL\_PROB $\geq$ 0.75), \Med\ demonstrates excellent recovery performance, achieving a rate of 98.3\% for sources with BI\_CIV $>$ 0, but only 56.1\% for those with BI\_CIV $=$ 0 and AI\_CIV $>$ 0. These results suggest that while the model is highly effective and comparable to existing automated methods in identifying \BALQSOs\ overall, significant discrepancies persist for sources with weak and narrow BAL troughs. Such troughs are particularly sensitive to  spectral preprocessing (e.g., interpolation and smoothing) and to spectral resolution (e.g., the compression of the probability vector labels). In future work, targeted improvements in these two aspects are expected to significantly enhance the detection of narrow and weak absorption. Furthermore, \Med\ newly identified 25,123 \BALQSOs, whose composite spectra exhibit prominent absorption troughs, further validating the model's capability to detect previously unrecognized BAL features.

This sample can be used to systematic investigate the cosmic evolutions of BAL properties (e.g., \citealt{2022_Bischetti_Natur.605..244B,2023_Bischetti_ApJ_952}), as traced by \ion{C}{4} BAL troughs, as well as the dependence of the BAL fraction with quasar nuclear properties (e.g., continuum slope, luminosity, black hole mass and accretion rate) (e.g., \citealt{2006_Trump_ApJS..165....1T,2007_Ganguly_ApJ_665,2014_Zhang_ApJ_786}). In addition, a subset comprising 20.7\% of our sample with at least two-epoch observations provides a valuable opportunity to investigate the BAL trough variability (e.g., \citealt{2013_Filiz_ApJ...777..168F,2015_Zhang_ApJ...803...58Z,2017_He_ApJS_229}). Monitoring changes in their strength, equivalent width, and velocity structure enables probing of the physical mechanisms governing outflows, such as acceleration, deceleration, or transverse motion. 
Notably, redshifted BAL troughs account for up to 8.8\% of all BAL troughs (10.9\% of the sources), representing the first large-scale, systematic survey specifically targeting these rare features.
Previous studies indicate that these redshifted BALs may originate either from rotating outflows close to the central black hole or from infalling (i.e., accreting) gas (e.g., \citealt{2013_Hall_MNRAS.434..222H,2016ApJ...829...96S,2019_Zhou_Natur.573...83Z,2021ApJ...916...86L}),
thus providing important clues to the gas kinematics near AGNs. 
Therefore, these systems offer valuable diagnostics for investigating black hole accretion processes.
A comprehensive multi-wavelength analysis of BAL troughs—including Balmer lines as well as metastable \ion{He}{1} and optical \ion{Fe}{2} and \ion{Mg}{2} in both optical and NIR bands—will be crucial for revealing the microphysical mechanisms that regulate black hole feeding.


In conclusion, the \Med\ framework represents a significant advancement in \BALQSO\ research. It simultaneously identifies \BALQSOs\ and measures their BAL troughs, while enhancing the efficiency of fitting unabsorbed spectra by leveraging the predicted properties of BAL troughs. Additionally, \Med\ can select troughs with varying confidence levels by setting different \PROB\ thresholds, thereby providing robust support for specific investigations. With its demonstrated reliability in BAL trough detection and characterization, \Med\ provides astronomers with a powerful tool for analyzing large quasar spectral datasets from current and future surveys.

\begin{acknowledgments}
This work is supported by the National Natural Science Foundation of China (Grant No. 12573009, 12173026 and 12141302), the Innovation Program of Shanghai Municipal Education Commission (Grant No. 2025GDZKZD04), and the scientific research grants from the China Manned Space Project (Grant No. CMS-CSST-2025-A06, CMS-CSST-2025-A07).
S.H.Z. acknowledges the support from the Program for Professor of Special Appointment (Eastern Scholar) at Shanghai Institutions of Higher Learning and the Shuguang Program (23SG39) of Shanghai Education Development Foundation and Shanghai Municipal Education Commission. 
This work makes use of data from SDSS-IV. Funding for SDSS has been provided by the Alfred P. Sloan Foundation and Participating Institutions. Additional funding toward SDSS-IV has been provided by the U.S. Department of Energy Office of Science. SDSS-IV acknowledges support and resources from the Center for High-Performance Computing at the University of Utah. The SDSS website is \url{www.sdss.org}.
\end{acknowledgments}

\bibliography{refs}{}
\bibliographystyle{aasjournalv7}

\end{CJK}
\end{document}